\documentclass[conference]{IEEEtran}
\IEEEoverridecommandlockouts
\usepackage{amsmath,amssymb,amsfonts}
\usepackage{algorithmic}
\usepackage{graphicx}
\usepackage{textcomp}
\usepackage{xcolor}
\usepackage{amsthm}
\usepackage{subfig}
\usepackage{booktabs} 
\usepackage{helvet}
\usepackage{courier}
\usepackage{graphicx}
\usepackage{stfloats}
\usepackage[all]{xy}
\usepackage{MnSymbol}
\usepackage{amsmath}
\usepackage{multirow}
\usepackage{stackengine}
\usepackage{fmtcount}
\usepackage[linesnumbered,ruled]{algorithm2e}
\usepackage[utf8]{inputenc}
\usepackage[english]{babel}
\usepackage{url}

\def\BibTeX{{\rm B\kern-.05em{\sc i\kern-.025em b}\kern-.08em
   T\kern-.1667em\lower.7ex\hbox{E}\kern-.125emX}}
\begin{document}

\title{DARKMENTION: A Deployed System to Predict Enterprise-Targeted External Cyberattacks}

\author{
	\IEEEauthorblockN{Mohammed Almukaynizi$^{1}$, Ericsson Marin$^{1}$, Eric Nunes$^{1}$, Paulo Shakarian$^{1,2}$}
	Gerardo I. Simari$^{3,1}$, Dipsy Kapoor$^{4}$, Timothy Siedlecki$^{5}$\\
	\IEEEauthorblockA{$^{1}$Arizona State University \\
		$^{2}$Cyber Reconnaissance, Inc. \\
		$^{3}$Department of Computer Science and Engineering, Universidad Nacional del Sur (UNS)\\
		Institute for C.S. and Eng. (CONICET–UNS)\\
		$^{4}$University of Southern California\\
		$^{5}$Lockheed Martin Advanced Technology Laboratories\\
		\{malmukay, esmarin, enunes1, shak\}@asu.edu}
	 gis@cs.uns.edu.ar, dipsy@isi.edu, timothy.siedlecki@lmco.com
	}

\maketitle

\thispagestyle{plain}
\pagestyle{plain} 
\IEEEpubidadjcol

\begin{abstract}
Recent incidents of data breaches call for organizations to proactively identify cyber attacks on their systems. Darkweb/Deepweb (D2web) forums and marketplaces provide environments where hackers anonymously discuss existing vulnerabilities and commercialize malicious software to exploit those vulnerabilities. These platforms offer security practitioners a threat intelligence environment that allows to mine for patterns related to organization-targeted cyber attacks. In this paper, we describe a system (called DARKMENTION) that learns association rules correlating indicators of attacks from D2web to real-world cyber incidents. Using the learned rules, DARKMENTION generates and submits warnings to a Security Operations Center (SOC) prior to attacks. Our goal was to design a system that automatically generates enterprise-targeted warnings that are timely, actionable, accurate, and transparent. We show that DARKMENTION meets our goal. In particular, we show that it outperforms baseline systems that attempt to generate warnings of cyber attacks related to two enterprises with an average increase in F1 score of about 45\% and 57\%. Additionally, DARKMENTION was deployed as part of a larger system that is built under a contract with the IARPA Cyber-attack Automated Unconventional Sensor Environment (CAUSE) program. It is actively producing warnings that precede attacks by an average of 3 days. 
     	
\end{abstract}

\section{Introduction}

With the widespread use of technology, cyber-security has become a concern for both commercial organizations and governments. With the recent incidents of data breaches at Equifax, Verizon, Gmail and others\footnote{https://www.identityforce.com/blog/2017-data-breaches}, organizations are looking at methods to proactively identify if they will be target of future attacks. A 2017 Verizon investigation report stated that 75\% of breaches were perpetrated by outsiders exploiting known vulnerabilities~\cite{verizon}. Monitoring the vulnerabilities that are of interest to malicious threat actors from the discussions on Darkweb/Deepweb (D2web) hacking sites is a key aspect of predicting cyber-attacks~\cite{Almukaynizi2017proactive}.

In this paper, we describe DARKMENTION, a system that identifies indicators of risks from unconventional sources of threat intelligence (D2web), monitors those sources in real-time to reason about the likelihood of future threats, generates warnings, and submits them to the Security Operations Center.

We use concepts from causal reasoning~\cite{suppes1970probabilistic,kleinberg2009temporal} and logic programming (in particular, the concepts of Point Frequent Function (\textit{pfr}) from APT-logic~\cite{Shakarian:2011,Shakarian:2012,stanton2015mining}) to learn association rules. An example of the rules we sought to learn is ``if certain D2web activity is observed in a given time-point, then there will be an $x$ number of attacks of type $y$, targeting organization $o$ in exactly $\Delta t$ time-points, with probability $p$”. Our data is obtained from a commercially available API, maintained by a cyber-threat intelligence firm (called CYR3CON\footnote{https://cyr3con.ai}), and from over 500 historical records of real-world targeted cyber incidents. Those incidents are recorded from the logs of two large enterprises participating to the IARPA Cyber-attack Automated Unconventional Sensor Environment (CAUSE) program\footnote{https://www.iarpa.gov/index.php/research-programs/cause}.

Throughout the paper, we illustrate the viability of DARKMENTION as a tool that addresses problems directly related to situational awareness, resource allocation, and countermeasure prioritization. In particular, we show that DARKMENTION produces warnings that are,
\begin{itemize}
	\item \textbf{timely}: indicates the exact time-point in which a predicted attack will occur,
	\item \textbf{actionable}: provides metadata/warning details, i.e., the target enterprise, type of attack, volume, and the software vulnerabilities/threat actor identified from the D2web discussions, 
	\item \textbf{accurate}: predicted unseen real-world attacks with an average increase in F1 of over 45\% for one enterprise and 57\% for the other, and
	\item \textbf{transparent}: allows analysts to easily trace the warnings back to the rules that were triggered, discussions that fired the rules, etc.
\end{itemize}

\vspace{0.5 em}
\section{Dataset Description}

\subsection{D2web Crawling Infrastructure}
Darkweb refers to the portion of the internet that is not indexed by search engines and cannot be accessed by standard browsers. Specialized browsers like Tor\footnote{See the Tor Project's official website (https://www.torproject.org/).} are required to access darkweb sites. We retrieve information from both \textit{marketplaces}: where users advertise to sell information regarding vulnerabilities or exploits, and \textit{forums}: that provide discussions on discovered vulnerabilities among others. 

We summarize the D2web crawling infrastructure that CYR3CON maintains\textemdash originally introduced in~\cite{Nunes:2016}. Customized lightweight crawlers and parsers were built for each site to collect and extract data. At the time of writing this paper, data is collected from more than 400 platforms (forums and marketplaces). To ensure collection of cybersecurity relevant data, machine learning models are used to filter data related to drugs, weapons and other discussions irrelevant to cybersecurity. \smallskip

\subsection{Data Pre-processing} 
\noindent\textbf{CVE-CPE mapping:} Common Vulnerability Enumeration (CVE) is a unique identifier assigned to each software vulnerability reported in the National Vulnerability Database (NVD~\cite{nvd}). Common Platform Enumeration (CPE) is a list of software/hardware products that are vulnerable to a given CVE. CPE data can be obtained from the NVD. We query the database using API calls to look for postings with software vulnerability mentions (in terms of CVE number). Regular expressions\footnote{https://cve.mitre.org/cve/identifiers/syntaxchange.html} are used to identify CVE mentions. 
We map each CVE to pre-identified groups of CPEs\footnote{We cluster CPEs together based on common vendors/products. We identified over 100 groups of CPEs, e.g., \textit{Microsoft-Office}, \textit{Apache-Tomcat}, and \textit{Intel}. However, only 33 were used as preconditions in the rules.} and nation-state threat actors who are known to leverage that CVE as part of their attack tactics\footnote{We have encoded a list of threat actors along with vulnerabilities they favor by manually analyzing cyberthreat reports that were recently published by cybersecurity companies, e.g., \url{https://media.kaspersky.com/en/business-security/enterprise/KL_Report_Exploits_in_2016_final.pdf}.}. Perhaps among the well-known threat actors is the North Korean group \textit{HIDDEN COBRA}, which was recently identified to account for an increasing number of cyberattacks to US targets\footnote{https://www.us-cert.gov/ncas/alerts/TA17-164A}. These CPE and nation-state actor mappings are used as pre-conditions during the rule-learning phase that is discussed in Sectiona~\ref{dep} and \ref{apt}.

\subsection{Enterprise-Relevant External Threats} 
To construct rules and evaluate the performance of the learned model, we use data from historical records of attack attempts that are recorded from the logs of two enterprises participating in the IARPA CAUSE program\footnote{https://www.iarpa.gov/index.php/research-programs/cause.}. One of the two enterprises is a defense industrial base (referred to Armstrong) while the other is a financial services organization (referred to Dexter). The database is distributed to the performers participating in a contest led by IARPA in increments, once every few months. Each data point is a record of a detected deliberate malicious attempt to gain unauthorized access, alter or destroy data, or interrupt services or resources in the environment of the participating organizations. Those malicious attempts were detected in uncontrolled environment, and by different security defense commercial products such as anti-virus, intrusion detection systems, and hardware controls. Each ground truth (GT) record includes: \textit{ID, Format Version, Reported Time, Occurrence Time, Event Type, and Target Industry}\footnote{We intentionally skip some details about other fields of the GT records due to the limitation in space and irrelevance to the scope of this paper.}. The types of attacks included in the GT dataset are:

\begin{itemize}
 \item \textbf{Malicious Email (M-E).} A malicious attempt is identified as a Malicious Email event if an email is received by the organization, and it either contains a malicious email attachment, or a link (embedded URL or IP address) to a known malicious destination.
  
 \item \textbf{Malicious Destination (M-D).} A malicious attempt is identified as a visit to a Malicious Destination if the visited URL or IP address hosts malicious content.
  
 \item \textbf{Endpoint Malware (E-M).} A Malware on Endpoint event is identified if malware is discovered on an endpoint device. This includes, but not limited to, ransomware, spyware, and adware.
\end{itemize}

Other events related to insider threats are out of the scope of the current phase of the deployed system. A summary of the time periods and the number of records for each attack type is provided in Section~\ref{res}.

\vspace{0.5 em}
\section{Deployed System}
\label{dep}

\begin{figure}[!h]
\centering
\includegraphics[scale=0.27,keepaspectratio]{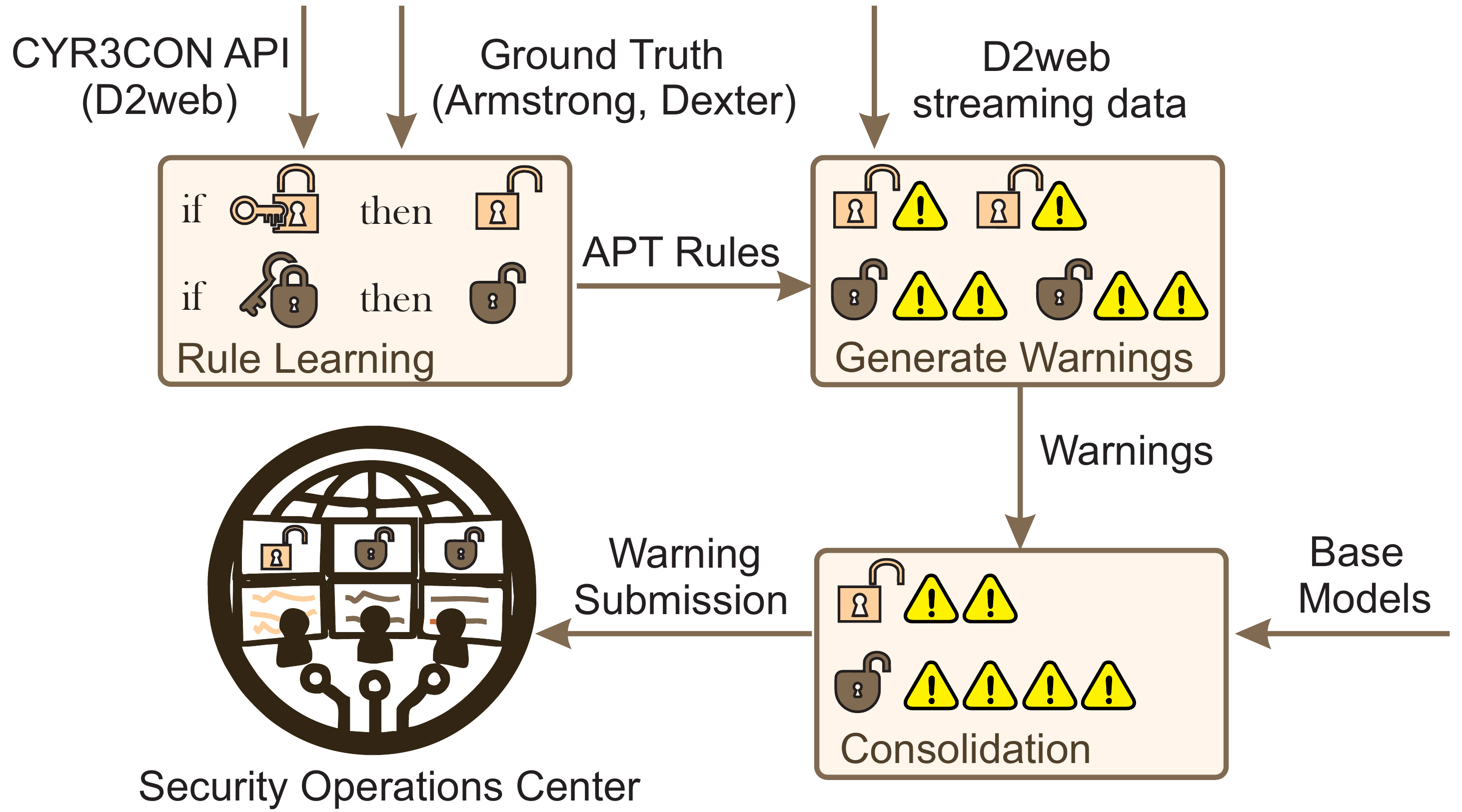}
 \caption{The Deployed System.}
\label{fig:Flow}
\end{figure}

In this section we provide an overview of DARKMENTION. Figure~\ref{fig:Flow} provides a graphical illustration of the system's workflow. Our system comprises three main components, each was designed to serve a set of tasks. There are three reasons for having this particular design: (1) there are other base models that were implemented using the same GT data, (2) this design would result in minimal changes in the roles of the subteams/members, and (3) model ensembles can improve the overall predictions of the model generated warnings~\cite{montgomery2012improving}.\smallskip

\begin{figure}[!h]
	\centering
	\subfloat[Armstrong]{\includegraphics[scale=0.16,keepaspectratio]{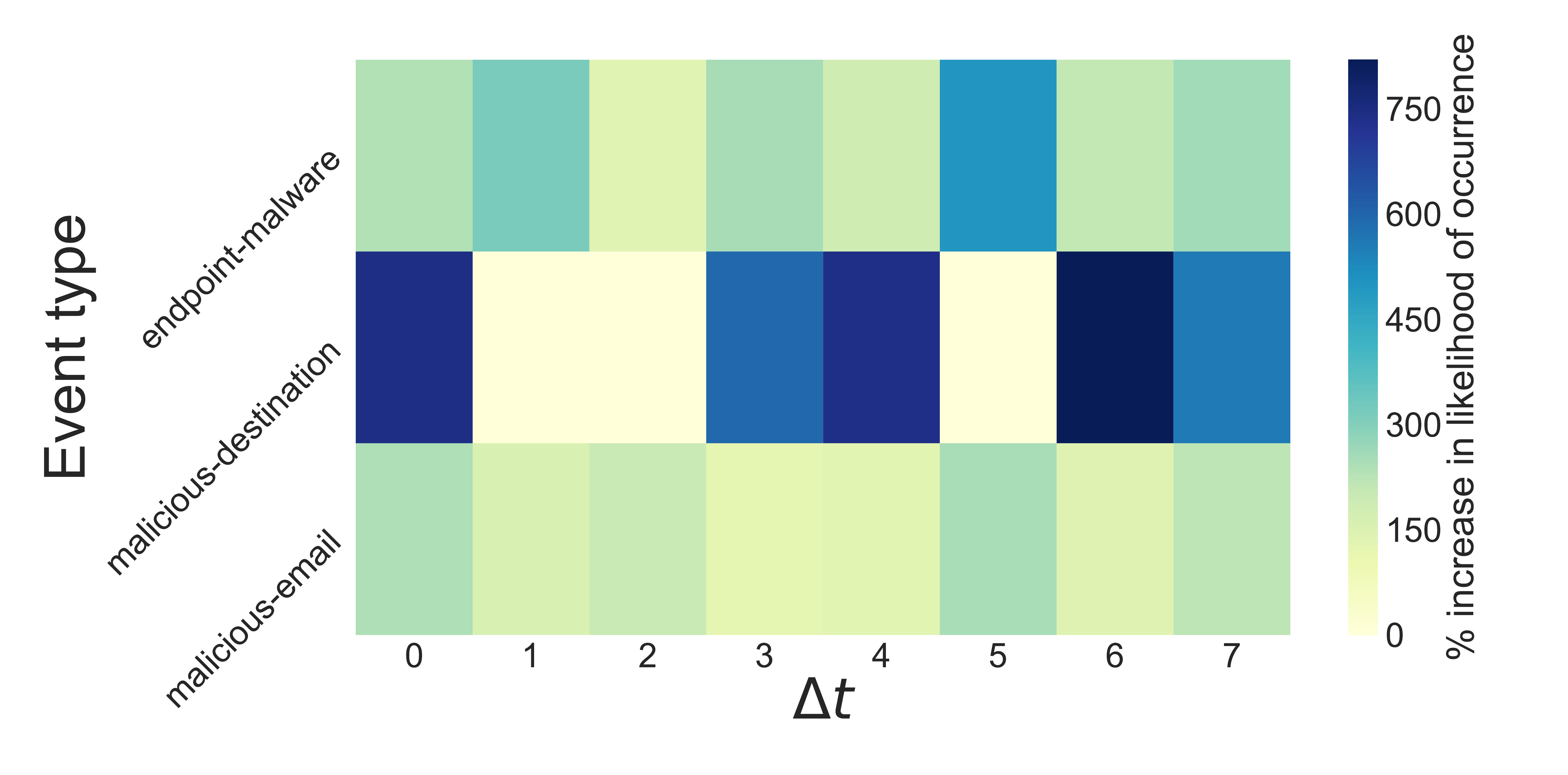}}
	\qquad
	\subfloat[Dexter]{\includegraphics[scale=0.153,keepaspectratio]{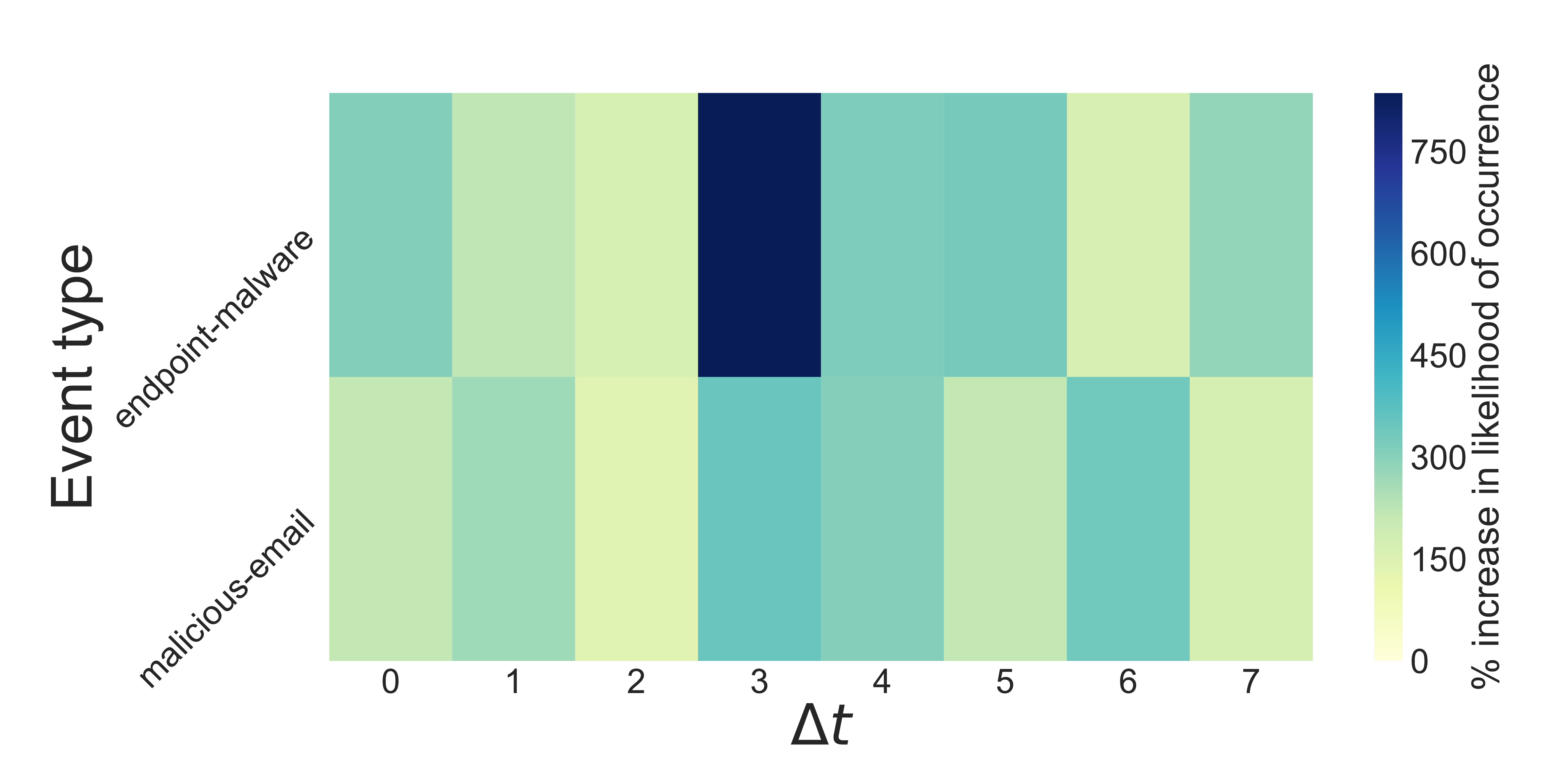}}
	\caption{The percentage increase in likelihood of occurrence for the rules that are identified to be useful.}
	\label{fig:heatmap_dexter_timely2}
	\vspace{-1em}
\end{figure}

\noindent\textbf{Rule learning:} The first component is responsible for learning rules, i.e., APT rules with \textit{pfr} function~\cite{Shakarian:2011,Shakarian:2012,stanton2015mining}. The inputs to the rule learner are: (1) the CPE-groups/actors that were identified from the D2web discussions and the time-point in which they are mentioned, and (2) the GT events with their types and time-points in which they were observed. The learner follows the technical approach discussed in Section~\ref{apt} to generate APT rules. The output of this component is a set of APT rules that are determined to be useful, i.e., exceeding some thresholds on probability of occurrence and some support count (frequency with which a rule is satisfied in the historical data). Such capability aids in producing \textit{accurate} and \textit{timely} warnings. Figure~\ref{fig:heatmap_dexter_timely2} shows the percentage increase in the likelihood of occurrence of attack events per days following D2web discussions for the rules that were identified as useful. We note that the reported records of Malicious Destination for Dexter only cover a time period that ends before the testing time period starts. Therefore, no rules were produced for Malicious Destination. Additionally, the system did not learn useful rules relating to Armstrong's Malicious Destination events when $\Delta t$ is 1, 2, or 5.
\vspace{-0.75em}

\noindent\textbf{Generating warnings:} The second component is responsible for generating warnings using the APT rules. This component is executed daily by first acquiring all CVEs mentioned in the last 24 hours within the D2web streaming data. The CPE groups/nation-state actors for these mentioned CVEs are then obtained. Next, based on the APT-rules, the model tries to match the CPE/nation-state actor mappings to a particular rule. If a match exists, the model predicts if and when an attack exploiting the vulnerabilities will occur by generating warnings. The warning fields are populated using the information contained in the rule, such as the probability, event type, and target organization. Such details help in producing \textit{actionable} warnings, i.e., warnings that provide metadata/details including the CVEs/tactics, industry, volume of discussions, etc. Section~\ref{rule-learning} provides further details about the way warnings are generated from rules.\smallskip

\begin{figure}[!h]
	\vspace{-0.5em}
	\centering
	\includegraphics[scale=0.75,keepaspectratio]{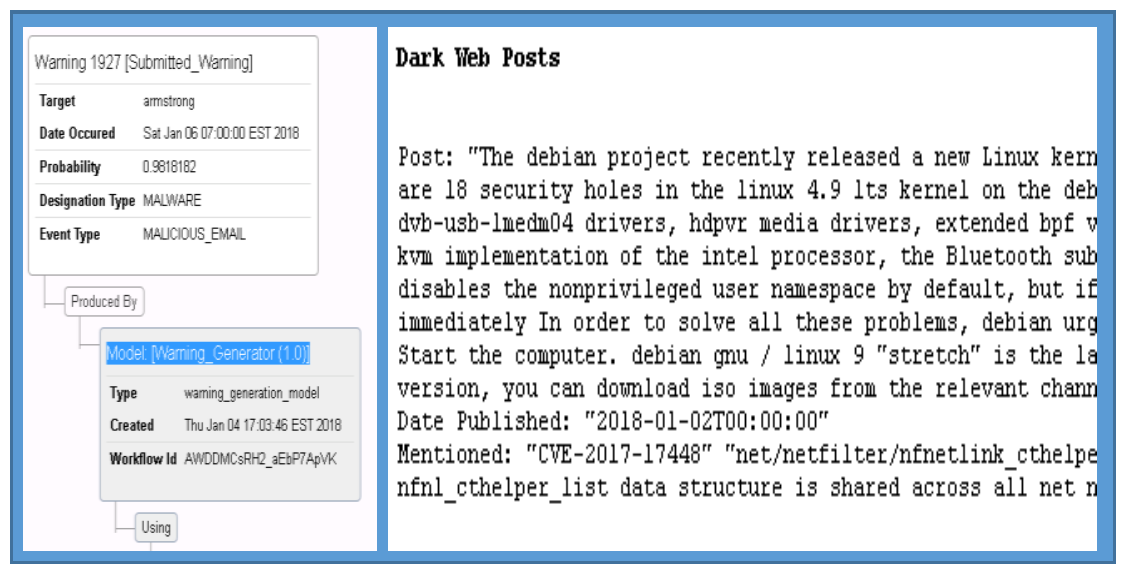}
	\caption{Two screenshots from DARKMENTION.}
	\label{fig:transparent}
\vspace{-0.5em}
\end{figure}

\noindent\textbf{Model fusion:} The final component deals with consolidation. It fuses warnings from various heterogeneous models (including DARKMENTION), populates any missing warning fields according to the program requirement, and generates the final version of each warning. This completed warning is submitted to the Security Operations Center. Each warning submitted is available to view and drill down into using a Web UI and Audit Trail analysis capability. This audit trail goes from the submitted warning all the way through model fusion, the individual models, and each individual model's raw data used to generate a warning. In the case of DARKMENTION, this would include the D2web postings/items with the CVEs mentioned highlighted. This capability makes the warnings that DARKMENTION produces \textit{transparent} warnings, i.e., allows analysts to easily trace the warnings back to the rules that were triggered, discussions that fired the rules, etc. Figure~\ref{fig:transparent} shows two screenshots taken from the system.

\section{Annotated Probabilistic Temporal Logic (APT-logic)}
\label{apt}

We will define here the syntax and semantics of APT-logic programs (set of APT-logic rules) applied to our domain, which is built upon the work of Shakarian et al. in \cite{Shakarian:2012}.  

\subsection{Syntax}

\vspace{-0.25em}
\noindent\textbf{Herbrand base.} We use $B_\mathcal{L}$ to denote the Herbrand base (finite set of ground atoms) of a first order logical language $\mathcal{L}$.  Then, we divide $B_\mathcal{L}$ into two disjoint sets: $B_{\mathcal{L}\{conditions\}}$ and $B_{\mathcal{L}\{actions\}}$, so that $B_\mathcal{L} \equiv B_{\mathcal{L}\{conditions\}} \cup \,\, B_{\mathcal{L}\{actions\}}$.  $B_{\mathcal{L}\{conditions\}}$ comprehends the atoms allowed only in the premise of APT rules, representing \textit{conditions} or users' actions performed on D2web websites, e.g., $\mathit{mention\_on(forum{\_1}, debian)}$. On the other hand, $B_{\mathcal{L}\{actions\}}$ comprehends the atoms allowed only in the conclusion of APT rules, representing \textit{actions} or malicious activities reported by Armstrong or Dexter organizations in their own facilities, e.g., $\mathit{attack(armstrong, malicious-email, x)}$.\smallskip 

\noindent\textbf{Formulas.} Complex sentences (formulas) are constructed recursively from atoms, using parentheses and the logical connectives: ($\neg$ negation, $\vee$ disjunction, $\wedge$ conjunction). However, we note that all formulas in this paper are single atoms.\smallskip 

\noindent\textbf{Time formulas.} If $F$ is a formula, $t$ is a time point, then $F_t$ is a time formula which states that $F$ is true at time $t$.\smallskip 

\noindent\textbf{Probabilistic time formulas.} If $\phi$ is a time formula and $[l,u]$ is a probability interval $\subseteq [0,1]$, then $\phi:[l,u]$ is a probabilistic time formula \textit{(ptf)}.  Intuitively, $\phi:[l,u]$ says $\phi$ is true with a probability in $[l,u]$, or using the complete notation, $F_t:[l,u]$ says $F$ is true at time $t$ with a probability in $[l,u]$.\smallskip

\noindent\textbf{APT rules.} Suppose condition $F$ and action $G$ are formulas, $t$ is a natural number, $[l,u]$ is a probability interval and $fr \in \mathcal{F}$ is a frequency function symbol that we will define later.  Then $F\overset{fr}{\rightlsquigarrow}G: [t,l,u]$ is an APT (Annotated Probabilistic Temporal) rule, which informally saying, computes the probability that $G$ is true in exactly $\Delta t$ time units after $F$ becomes true.  For instance, the APT rule below informs that the probability of Armstrong company being attacked by a malicious-email, in exactly 3 time units after users mention \textit{``debian''} on \textit{forums{\_1}}, is between 40\% and 50\%.

\vspace{5pt}

\scriptsize

\noindent mention{\_on(set\_forum{\_1},debian)} $\,\overset{pfr}{\rightlsquigarrow}\,$ attack(armstrong, malicious-email)$\,\,:\,\,$[3,0.4,0.5] 

\normalsize

\subsection{Semantics}

\noindent\textbf{World.} In general, a world is any set of ground atoms that belong to $B_\mathcal{L}$.  However, due to our constraint that separates atoms into $B_{\mathcal{L}\{conditions\}}$ and $B_{\mathcal{L}\{actions\}}$, not all possible worlds are allowed in our APT rules. Strictly, one atom belonging to $B_{\mathcal{L}\{conditions\}}$ and one atom belonging to $B_{\mathcal{L}\{actions\}}$ must be present in an allowable world for our \textit{pfr} rules.\smallskip

\noindent\textbf{Thread.} A thread is a series of worlds that model the domain over time, where each world corresponds to a discrete time-point in $\mathcal{T} = \{1,..., t_{max}\}$.  $Th(i)$ specifies that according to the thread $Th$, the world at time $i$ will be $Th(i)$.   Given a thread $Th$ and a time formula $\phi$, we say $Th$ satisfies $\phi$ (denoted $Th \models \phi$) iff:

\begin{itemize}
  \item If $\phi \equiv F_t$ for some ground time formula $F_t$, then $Th(t)$ satisfies $F$;
	\item If $\phi \equiv \neg \rho$ for some ground time formula $\rho$, then $Th$ does not satisfy $\rho$;
	\item If $\phi \equiv \rho_1 \wedge \rho_2$ for some ground time formulas $\rho_1$ and $\rho_2$, then $Th$ satisfies $\rho_1$ and $Th$ satisfies $\rho_2$;
	\item If $\phi \equiv \rho_1 \vee \rho_2$ for some ground time formulas $\rho_1$ and $\rho_2$, then $Th$ satisfies $\rho_1$ or $Th$ satisfies $\rho_2$;	
\end{itemize}

\noindent\textbf{Frequency functions.} A frequency function represents temporal relationships within a thread, checking how often a world satisfying formula $F$ is followed by a world satisfying formula $G$.  Formally, a frequency function $fr$ belonging to $\mathcal{F}$ maps quadruples of the form $(Th,F,G, t)$ to [0,1] of real numbers. Among the possible ones proposed in \cite{Shakarian:2011}, we investigate here alternative definitions for \textit{Point Frequency Function (pfr)}, which specifies how frequently the action $G$ follows the condition $F$ in ``exactly'' $\Delta t$ time points. To support ongoing security operations we need to relax the original assumption of a finite time horizon $t_{max}$ in~\cite{Shakarian:2011,Shakarian:2012}. We introduce here a different but equivalent formulation for \textit{pfr} that does not rely on a finite time horizon. To accomplish that, we first need to define how a \textit{ptf} can be satisfied in our model.  If we consider the \textit{ptf} $F_t:[l,u]$, and some $A' \in A$, where $A$ is the set of all \textit{ptf}'s satisfied by our thread $Th$, we say $Th \models F_t:[l,u]$ iff:

\vspace{3pt}

\begin{itemize}
  \item If $F = a$ for some ground $a$, then $\exists \, a_t:[l',u'] \in A   \,\, s.t.   \,\, [l',u'] \sqsupseteq [l,u]$;
  \vspace{3pt}
  \item If $F_t:[l,u] = \neg F'_t:[l,u]$ for some ground formula $F'$, then $Th \models F'_t:[1-u,1-l]$;
  \vspace{3pt}
  \item If $F_t:[l,u] = F'_t:[l,u] \wedge F''_t:[l,u]$ for some ground formulas $F'$ and $F''$, then $Th \models F'_t:[l,u]$ and $Th \models F''_t:[l,u]$;
  \vspace{3pt}
  \item If $F_t:[l,u] = F'_t:[l,u] \vee F''_t:[l,u]$ for some ground formulas $F'$ and $F''$, then $Th \models F'_t:[l,u]$ or $Th \models F''_t:[l,u]$;
\end{itemize}

\vspace{3pt}

We are ready to show the new formulation of \textit{pfr} in Equation~\ref{eq1}, which is equivalent to the original one proposed in \cite{Shakarian:2011} when $t_{max}$ comprises the whole thread $Th$ (all time points):

\vspace{-3pt}

\begin{equation}\label{eq1}
\hspace{-168pt} pfr(Th,F,G,\Delta t) =
\end{equation}

\vspace{-20pt}
\begin{multline*}
\hspace{-12pt} \left[ \frac{\sum\limits_{t| Th \models F_t:[l,u] \wedge Th \models G_{t+\Delta t}:[l',u']}l'}{\sum\limits_{t| Th \models F_t:[l,u]}u}, \frac{\sum\limits_{t| Th \models F_t:[l,u] \wedge Th \models G_{t+\Delta t}:[l',u']}u'}{\sum\limits_{t: Th \models F_t:[l,u]}l} \right]  
\end{multline*}

\vspace{3pt}

\noindent\textbf{Satisfaction of APT rules and programs.} We say $Th$ satisfies an APT Rule $F\overset{pfr}{\rightlsquigarrow}G: [\Delta t,l,u]$ (denoted $Th \models F\overset{pfr}{\rightlsquigarrow}G: [\Delta t,l,u]$) iff: 

\vspace{-0.5em}

\begin{equation}\label{eq4}
\centering
pfr(Th,F,G,\Delta t) \subseteq [l,u]
\end{equation}

\subsection{Rules and warnings}
\label{rule-learning}

\noindent\textbf{Probability intervals.} We derive the probability intervals related to all pairs $[l,u]$ specified in this paper using the standard deviation of the corresponding point probability in a binomial distribution, i.e., $std(p) = \frac{n * p * (1-p)}{n}$~\cite{Wadsworth:1960}, where $n$ is the number of events that produced the probability $p$.\smallskip

\noindent\textbf{APT programs.} Our algorithms only adds to the logic programs the \textit{pfr} rules with lower bounds exceeding the prior probability of the rule's head happening at any random time point.\smallskip

\noindent\textbf{Warnings generation.} The problem is to identify whether a triggered rule should generate warnings, and the number of warnings to generate. When there is no triggered rules on a given day, no warnings are generated. When two rules are triggered on the same day, both predict the same attack type is going to happen on the same day (i.e., same $\Delta t$), and one predicts $x$ number of attacks while the other predicts $y$ number of attacks, then they will generate $x + y$ warnings if both are qualified to generate warnings. A \textit{pfr} $r \in R$ is qualified to generate $x$ number of warnings if (1) the rule is triggered, and (2) there is no other triggered rule $r' \in R$ on the same day and same rule head and $\Delta t$ as $r$'s but has higher point probability than $r$'s.\smallskip

\section{Experimental Results}
\vspace{-0.20em}
In this section, we provide evidence of the viability of our approach through a series of experiments. The warnings that are submitted by our system are evaluated by the Security Operations Centers (SOCs) on a monthly basis. However, we internally evaluated DARKMENTION since the external evaluations are aggregated for all models, and DARKMENTION was operationally deployed only after the time periods those reports cover.

\subsection{Experimental Settings}
\vspace{-0.5em}
We perform evaluations on the warnings targeting Armstrong that were submitted during July, August, and September of 2017. The results are aggregated per months for the experiments on Armstrong data while aggregated on periods of 7 days for Dexter. The latter starts from July 1 to July 28, 2016. These time windows differ because the Armstrong dataset covers a longer period of time as compared to the period covered by Dexter, and there is no more GT data about Dexter that is going to be provided or evaluated by the program. The reported records of Malicious Destination for Dexter only cover a time period that ends before the testing time period starts, hence they are not evaluated.

\subsection{Evaluation Metrics}
\vspace{-0.5em}
To evaluate the accuracy of our system, we use three metrics: recall, which corresponds to the fraction of GT events that have matching warnings from the total number of GT events; precision, which is the fraction of warnings that have matching GT events from the total number of the generated warnings; and F1, which is the harmonic mean of recall and precision.\smallskip 

%
%
%
%
%

\noindent\textbf{Matching warnings and GT events.} DARKMENTION predicts the exact number of attacks on each day.  Therefore, the matching problem is to find whether a warning $w$ earns credit for predicting a GT attack event $g$. If $w$ predicts an attack with different type than $g$\textquotesingle s, or $w$ predicts an attack on a different day than the occurrence day of $g$, then they do not match. Otherwise, they may or may not match based on whether or not $w$ or $g$ have already been paired up with another GT event or a warning, respectively. 

To join together warnings with GT events in parings while ensuring that resulting pairings are mutually exclusive, we use the Hungarian assignment algorithm \cite{munkres1957algorithms}. Intuitively, the algorithm takes as input an $n*n$ matrix representation of ($-1 * $ lead-time). Lead-time is the time between warnings and GT events that are qualified to match. Then, the algorithm returns a solution $S$ that maximizes the total lead-time\footnote{Lead-time is a metric set by the program to evaluate the performers, but we ignore it in this evaluation.}. Here, $S$ is a set of pairs, each maps a warnings to a GT event such that the pairs are guaranteed to be mutually exclusive\footnote{We add dummy rows (warnings) or columns (GT events) with lead-times of 1 to make the matrix square, then remove from the solution the pairs where the lead-time is 1.}. We store in the database the pairs that are returned by the algorithm.

\vspace{-0.25em}

\subsection{Results}
\label{res}
\vspace{-0.5em}
We found that our system outperforms a baseline system that randomly generates $x$ number of warnings on each day such that each value of $x$ has a chance proportional to its frequency of occurrence in the historical data. We repeat the baseline for 100 runs and take the average of each metric. In the real-time deployment of DARKMENTION, human experts can evaluate the warnings by leveraging the other capabilities of the system, i.e., \textit{transparency} and \textit{actionable} through a Web UI dashboard. However, in those experiments any triggered rule is counted, which is not necessarily important given other details. That said, our system scored significantly higher than the baseline system as shown in Table~\ref{Dexter-pfr_exact}.

\begin{table*}[]
	\tabcolsep=0.13cm
\small
\centering
\caption{Evaluation results. * A simple baseline model that generates $x$ number of warnings on each day based on the prior probability of each possible value of $x$ that was seen in the training data.}
\label{Dexter-pfr_exact}
\begin{tabular}{|l|l|c|c|c|c|c|c|c|c|c|c|l|}
	\hline
	\multirow{2}{*}{Dataset}   & \multirow{2}{*}{type} & \multirow{2}{*}{Testing starts} & \multirow{2}{*}{\#GT-events} & \multicolumn{4}{c|}{DARKMENTION}                                                                                 & \multicolumn{4}{c|}{Baseline* (average of 100 runs)}                                                             & \multirow{2}{*}{\begin{tabular}[c]{@{}l@{}}\%increase\\  in F1\end{tabular}} \\ \cline{5-12}
	&                       &                                 &                              & \#warnings & Precision                       & Recall                          & F1                              & \#warnings & Precision                       & Recall                          & F1                              &                                   \\ \hline\hline
	\multirow{9}{*}{Armstrong} &\multirow{3}{*}{M-E}                   & Jul-17                          & 24                           & 32         & 0.313                           & \textbf{0.417} & \textbf{0.357} & 11.759     & \textbf{0.417} & 0.205                           & 0.271                           & 32\%                              \\ \cline{3-13} 
	&                       & Aug-17                          & 11                           & 3          & \textbf{1.000} & 0.273                           & \textbf{0.429} & 11.966     & \textbf{0.289} & 0.315                           & 0.299                           & 43\%                              \\ \cline{3-13} 
	&                       & Sep-17                          & 13                           & 18         & 0.167                           & 0.231                           & 0.194                           & 12.793     & \textbf{0.249} & \textbf{0.249} & \textbf{0.247} & -21\%                             \\ \cline{2-13} 
	&\multirow{3}{*}{M-D}                  & Jul-17                          & 4                            & 12         & \textbf{0.167} & \textbf{0.500} & \textbf{0.250} & 3.534      & 0.099                           & 0.091                           & 0.090                           & 178\%                             \\ \cline{3-13} 
	&                       & Aug-17                          & 9                            & 23         & 0.174                           & \textbf{0.444} & \textbf{0.250} & 3.121      & \textbf{0.232} & 0.086                           & 0.120                           & 108\%                             \\ \cline{3-13} 
	&                       & Sep-17                          & 3                            & 10         & \textbf{0.100} & \textbf{0.333} & \textbf{0.154} & 2.948      & 0.071                           & 0.075                           & 0.068                           & 126\%                             \\ \cline{2-13} 
	& \multirow{3}{*}{E-M}                   & Jul-17                          & 14                           & 10         & 0.300                           & \textbf{0.214} & \textbf{0.250} & 8.552      & \textbf{0.326} & 0.200                           & 0.242                           & 3\%                               \\ \cline{3-13} 
	&                       & Aug-17                          & 18                           & 45         & 0.200                           & \textbf{0.500} & \textbf{0.286} & 9.155      & \textbf{0.324} & 0.168                           & 0.217                           & 32\%                              \\ \cline{3-13} 
	&                       & Sep-17                          & 17                           & 21         & \textbf{0.286} & \textbf{0.353} & \textbf{0.316} & 8.879      & 0.247                           & 0.127                           & 0.164                           & 93\%                              \\ \hline \hline
	\multirow{8}{*}{Dexter}    & \multirow{3}{*}{M-E}                   & 1-Jul-16                        & 2                            & 13         & 0.150                           & \textbf{1.000} & \textbf{0.267} & 2.720      & \textbf{0.157} & 0.205                           & 0.169                           & 58\%                              \\ \cline{3-13} 
	&                       & 8-Jul-16                        & 7                            & 10         & 0.500                           & \textbf{0.714} & \textbf{0.588} & 2.610      & \textbf{0.655} & 0.253                           & 0.348                           & 69\%                              \\ \cline{3-13} 
	&                       & 15-Jul-16                       & 9                            & 6          & 0.333                           & \textbf{0.222} & 0.267                           & 2.770      & \textbf{0.619} & 0.188                           & \textbf{0.276} & -3\%                              \\ \cline{3-13} 
	&                       & 22-Jul-16                       & 4                            & 2          & \textbf{0.500} & 0.250                           & 0.333                           & 3.050      & 0.469                           & \textbf{0.355} & \textbf{0.385} & -14\%                             \\ \cline{2-13} 
	& \multirow{3}{*}{E-M}                   & 1-Jul-16                        & 1                            & 2          & \textbf{0.500} & \textbf{1.000} & \textbf{0.667} & 1.790      & 0.189                           & 0.330                           & 0.226                           & 195\%                             \\ \cline{3-13} 
	&                       & 8-Jul-16                        & 3                            & 4          & \textbf{0.250} & \textbf{0.333} & \textbf{0.286} & 1.750      & 0.245                           & 0.167                           & 0.186                           & 54\%                              \\ \cline{3-13} 
	&                       & 15-Jul-16                       & 3                            & 1          & \textbf{1.000} & \textbf{0.333} & \textbf{0.500} & 1.740      & 0.281                           & 0.190                           & 0.217                           & 130\%                             \\ \cline{3-13} 
	&                       & 22-Jul-16                       & 4                            & 2          & \textbf{0.500} & \textbf{0.250} & \textbf{0.333} & 1.780      & 0.383                           & 0.208                           & 0.257                           & 30\%                              \\ \hline
\end{tabular}
\vspace{0.5 em}
\end{table*}

\vspace{-0.25em}
\section{Related Work.}
\vspace{-0.25em}
To the best of our knowledge, this paper is the first to present a deployed system that applies causal reasoning to predict enterprise-related external cyber threats. However, there exists a large body of related research with studies about malicious hacking community in D2web, and studies related to rule-learning methods for security applications.\smallskip

\noindent\textbf{Hacking community on D2web.}. While the hacking community in D2web sites has been widely investigated in the existing literature for applications such as analyzing the economics of D2web forums/markets \cite{motoyama2011analysis,allodi2017economic} and identifying future cyber-threats to mitigate risks \cite{Almukaynizi2017proactive,Tavabi2018iaai}, none of these studies identify threats related to specific corporations or identify when in the future the predicted events may occur. DARKMENTION specifically predicts enterprise-targeted attacks and the periods in which those threats are predicted.\smallskip

\noindent\textbf{Rule-learning methods for security applications.} A large body of work on understanding and modeling the behavior of threat actors and reasoning future risk levels to assist in the implementation of strategic defense has been proposed. While a growing number of studies along these lines have focused on applications related to understanding the behavior of terrorist and insurgent groups \cite{subrahmanian2013indian,stanton2015mining}, the cyber-security applications have not received much attention, except in the line of graph-based and attacker/defender game theoretical modeling~\cite{robertson2016data,brown2014addressing}. Our work differs from such theoretical approaches since we focus on practical details related to deployment, and evaluate our system with real-world data.\smallskip

\section{Conclusion}
We present DARKMENTION, a deployed system that produces warnings about cyber incidents that will likely occur in the future. Although the problem is difficult, our system proves to be useful as a tool that helps SOC teams to identify risks, potential sources of risk (vulnerabilities or threat actors) and context on which it builds its reasoning in a timely, actionable, accurate, and transparent manner. Our team was selected to progress to the second phase of the CAUSE program. 

\vspace{0.5 em}
\section*{Acknowledgment}
Some of the authors were supported by the Office of Naval Research (ONR) Neptune program, the ASU Global Security Initiative (GSI), and the National Council for Scientific and Technological Development (CNPq-Brazil).  Paulo Shakarian, Dipsy Kapoor, and Timothy Siedlecki are supported by the Office of the Director of National Intelligence (ODNI) and the Intelligence Advanced Research Projects Activity (IARPA) via the Air Force Research Laboratory (AFRL) contract number FA8750-16-C-0112. Gerardo Simari is also partially supported by Universidad Nacional del Sur (UNS) and CONICET, Argentina. The U.S. Government is authorized to reproduce and distribute reprints for Governmental purposes notwithstanding any copyright annotation thereon. Disclaimer: The views and conclusions contained herein are those of the authors and should not be interpreted as necessarily representing the official policies or endorsements, either expressed or implied, of ODNI, IARPA, AFRL, or the U.S. Government.

\vspace{0.5 em}
\bibliographystyle{IEEEtran} 
\bibliography{ISI-Darkmentions-camera-version}

\end{document}